# The cosmic ray luminosity of the nearby active galactic nuclei

L.G. Dedenko[1], D.A. Podgrudkov[1], T.M. Roganova[2], G.F. Fedorova[2]

[1] *Faculty of physics, M.V. Lomonosov Moscow State University, Moscow 119992, Leninskie Gory, Russia*
[2] *D.V. Skobeltsin Institute of Nuclear Physics, M.V. Lomonosov Moscow State University, Moscow 119992, Leninskie Gory, Russia*

## Abstract

The pointing directions of extensive air showers observed at the Pierre Auger Observatory were fitted within $\pm\ 3.1^{\circ}$ with positions of the nearby active galactic nuclei from the Véron-Cetty and P. Véron catalog. The cosmic ray luminosity of the active galactic nuclei which happened to be a source of the particular cosmic ray event constitutes a fraction $\sim 10^{-4}$ of the optical one if only cosmic ray particles with energies above $6\times 10^{19}$ eV are produced. If produced cosmic ray particles have a spectrum $dE/E^3$ up to ~100 GeV then the cosmic ray luminosity would be much higher than the optical one of the active galactic nuclei.

## 1. Introduction

At the very beginning of measurements in the southern semi-sphere the Pierre Auger Collaboration has reported that arrival directions of ultrahigh energy cosmic rays are not isotropic [1-3]. Moreover a remarkable correlations between the pointing directions of the Pierre Auger Observatory (PAO) events and positions of relatively nearby active galactic nuclei (AGN) from the Véron-Cetty and P. Véron (VCV) catalog [4] has been observed [1-3]. This profound discovery opens the new era of the cosmic ray astronomy [5,6]. It is a puzzle that in the northern semi-sphere such correlations between pointing directions of ultrahigh energy cosmic rays and positions of the AGNs from the VCV catalog have not been observed [7].

Our analysis of data of the Yakutsk array [8] which is also in the northern semi-sphere showed rather isotropic distribution of arrival directions of ultrahigh energy cosmic rays than a clustering in the super galactic plane. But later analysis of the Yakutsk data [9] has claimed the correlation with the AGN.

Searching for correlations between the pointing directions of ultrahigh energy cosmic rays and positions of some prominent objects in the sky allows to unveil the possible sources of these cosmic rays. Such correlations with the quasars [10], some objects in the supergalactic plane [11], the BL Lacertae [12] and Seifert galaxies [13] have been reported.

A suggestion [1-3] that the nearby active galactic nuclei (or other objects with the same spatial distribution) may be the possible sources of ultrahigh energy cosmic rays (the AGN hypothesis) is very attractive. However this suggestion as it was pointed out in Refs [14,15] has some problems. Namely, a profound deficit of ultrahigh energy cosmic ray events from the Virgo cluster was remarked [14,15]. It should be remembered that one of the paper has such subtitle as "All roads lead back to Virgo" [16]. In this paper we discuss some common requirements to the celestial objects which are supposed to be the sources of the ultrahigh energy cosmic rays. Namely, we estimate the cosmic ray luminosity of the nearby AGN which should be powerful enough to produce the PAO events.

## 2. Cosmic ray luminosity

First of all we have been searching for coincidences (within ± 3.1°) of the pointing directions of the ultrahigh energy cosmic rays observed [3] with the positions of the AGN from the VCV catalog [4]. If a pointing direction of a particular extensive air shower (EAS) coincides (within ± 3.1°) with positions of several AGN (e.g. with m objects) then each AGN (of this sample of m objects) was assigned a weight w (w=1/m). To estimate the cosmic ray luminosity we assumed that any particular sources are viewed only a fraction 1/3 of the total exposure time. We ignore also the absorption of the ultrahigh energy cosmic rays due to the Greizen-Zatsepin-Kuzmin (GZK) effect [17,18].

If the energy of 10 J is assigned to each ultrahigh energy cosmic ray particle then in case when sources emit isotropic cosmic rays the cosmic ray luminosity $LCR$ of any object at distance R from the Earth may be estimated by simple formula:

$$LCR = 1.27 \times 10^{30} (R/1Mpc)^2, \text{W}. \qquad (1)$$

It should be stressed that this is the luminosity of cosmic rays only with energies $E$ above 57 EeV (near 10 J). Another way to estimate the cosmic ray luminosity is to assume that one cosmic ray particle with the energy above 57 EeV hits each square kilometer on the sphere with a radius $R$=78.6 Mpc every hundred years. Then the power of losses is easily estimated as $1.64 \times 10^{35}$ W. Assuming that nearly ~300 objects are in the field of view of the PAO we arrive to the mean estimate of the cosmic ray luminosity $LCR$= $5.6 \times 10^{32}$ W.

These very simple assumptions allow us to construct the Table 1. The first 4 columns of the table show the number of the Auger events, galactic coordinates and the energy of an EAS.

The column 5 displays the celestial objects from [4] which were found to be within ± 3.1° from the    Table 1

| 1 | 2 | 3 | 4 | 5 | 6 | 7 | 8 | 9 | 10 | 11 |
|---|---|---|---|---|---|---|---|---|---|---|
|   | Showers | | | AGN | | | | | | |
| № | Glong, deg | Glat, deg | E, EeV | Name | Glong, deg | Glat, deg | R, Mpc | $L_0$, $10^{36}$ W | LCR, $10^{32}$ W | LR, $10^{35}$ W |
| 1 | 15.4 | 8.4 | 70 | | | | | | | |
| 2 | 309.2 | 27.6 | 84 | ESO 383-G18 | 312.8 | 28.1 | 54.6 | 3.30 | 3.56 | |
| 3 | 310.4 | 1.7 | 66 | 4U 1344-60 | 309.8 | 1.5 | 54.6 | 0.02 | 3.56 | |
| 4 | 392.3 | -17.0 | 83 | ESO 139-G12 | 332.4 | -14.7 | 74.4 | 5.64 | 6.08 | |
| 5 | 325.6 | 13.0 | 63 | IC 4518A | 326.1 | 14.0 | 67.2 | 0.04 | 5.39 | |
| 6 | 284.4 | -78.6 | 84 | NGC 424 | 283.2 | -78.3 | 46.2 | 1.46 | 2.55 | |
| 7 | 58.8 | -42.4 | 71 | | | | | | | |
| 8 | 307.2 | 14.1 | 58 | NGC 4945<br>ESO 269-G12 | 305.3<br>303.9 | 13.3<br>16.0 | 8.4<br>67.2 | 0.04<br>1.81 | 8.42<br>5.39 | 2.03 |
| 9 | 4.2 | -54.9 | 57 | | | | | | | |
| 10 | 48.8 | -28.7 | 59 | Zw 374.029<br>UGC 11630 | 50.5<br>47.5 | -26.0<br>-25.4 | 54.6<br>50.4 | 0.83<br>5.87 | 3.56<br>3.03 | |
| 11 | 256.3 | -10.3 | 84 | | | | | | | |
| 12 | 196.2 | -54.4 | 78 | SDSS J03349-0548 | 191.7 | -45.7 | 75.6 | 0.18 | 6.82 | |
| 13 | 332.4 | -16.5 | 59 | ESO 139-G12 | 332.4 | -14.7 | 71.4 | 5.64 | 6.08 | |
| 14 | 307.7 | 7.3 | 79 | IC 4200 | 305.8 | 10.8 | 54.6 | 2.69 | 3.56 | |
| 15 | 88.8 | -47.1 | 83 | NGC 7591 | 85.8 | -49.4 | 71.4 | 6.36 | 6.08 | |
| 16 | 189.4 | -45.7 | 69 | SDSS J03302-0532<br>NGC 1358<br>MARK 607 | 190.4<br>190.6<br>186.4 | -46.5<br>-45.6<br>-46.2 | 54.6<br>54.6<br>37.8 | 0.25<br>5.47<br>1.09 | 3.56<br>3.56<br>1.70 | |
| 17 | 308.8 | 17.2 | 69 | NGC 5244 | 311.5 | 16.2 | 33.6 | 1.65 | 1.35 | |
| 18 | 302.8 | 41.8 | 148 | | | | | | | |
| 19 | 63.5 | -40.2 | 58 | Q2207+0122 | 63.0 | -41.8 | 54.6 | 0.03 | 3.56 | |
| 20 | 308.6 | 19.2 | 70 | ESO 323-G77<br>NGC 5128 | 306.0<br>309.5 | 22.4<br>19.4 | 63.0<br>4.2 | 5.18<br>0.04 | 4.74<br>2.10 | 20.12 |
| 21 | 250.6 | 23.8 | 64 | ESO 565-G10<br>NCG 2989 | 253.9<br>253.0 | 21.7<br>26.0 | 63.0<br>54.6 | 4.15<br>1.44 | 4.74<br>3.56 | |
| 22 | 196.2 | -54.4 | 78 | MCG-02.09.040<br>NGC 1204 | 198.2<br>194.2 | -51.1<br>-55.5 | 63.0<br>58.8 | 2.65<br>2.20 | 4.74<br>4.12 | |
| 23 | 318.3 | 5.9 | 64 | | | | | | | |
| 24 | 12.1 | -49.0 | 90 | NGC 7135<br>IC 5135 | 10.1<br>10.0 | -50.6<br>-50.4 | 29.4<br>67.2 | 1.26<br>3.90 | 1.03<br>5.39 | |
| 25 | 338.2 | 54.1 | 71 | NGC 5506 | 339.2 | 53.8 | 29.4 | 0.47 | 1.03 | |
| 26 | 294.9 | 34.5 | 80 | | | | | | | |
| 27 | 234.8 | -7.7 | 69 | | | | | | | |

shower arrival direction. Galactic coordinates, a distance *R* from the Earth and an optical luminosity $L_0$ of the AGN found occupy next four columns. The cosmic ray luminosity *LCR* occupies the 10-th column and the Roentgen and gamma luminosity *LR* taken from [19-21] is shown for comparison in the last 11-th column.

With the help of this table and the VCV catalog [4] it is possible to make some analysis. First of all the well known anisotropy of the matter distribution within ~100 Mpc is illustrated by Fig. 1 where the curve 1 shows the lateral distribution of the AGN from [4] and curve 2 is an expected distribution in case of isotropy normalized in the 5-th bin to the curve 1. The total cosmic ray luminosity *LCRE57* of all sources inside a bin (with a width of 8.4 Mpc) divided by the number $N_i$ of the AGN inside the same bin is shown in Fig. 2. So it is assumed that all AGN inside a bin are capable to accelerate cosmic ray particles but only a few of them have been recorded by chance as sources. In agreement with [14,15] there are no cosmic ray sources in the interval 5 − 25 Mpc where the Virgo cluster is placed. Then a rather strange tendency of Figure increasing luminosity with a distance is displayed. If the AGN are standard sources then we should expect nearly constant luminosity. Fig. 3 illustrates this expectation for case of the optical luminosity of the AGN. As some kind of efficiency of the cosmic ray sources the ratio β of cosmic ray luminosity to the optical one is shown in Fig. 4. It is not easy to understand the growing efficiency with the distance. Besides, this efficiency is rather high. If we assume that the cosmic ray particles are produced with energy spectrum $dE/E^3$ up to energies ~100 GeV, then a value $β$ should be multiplied by a factor of ~$10^9$! It is much above the optical luminosity.

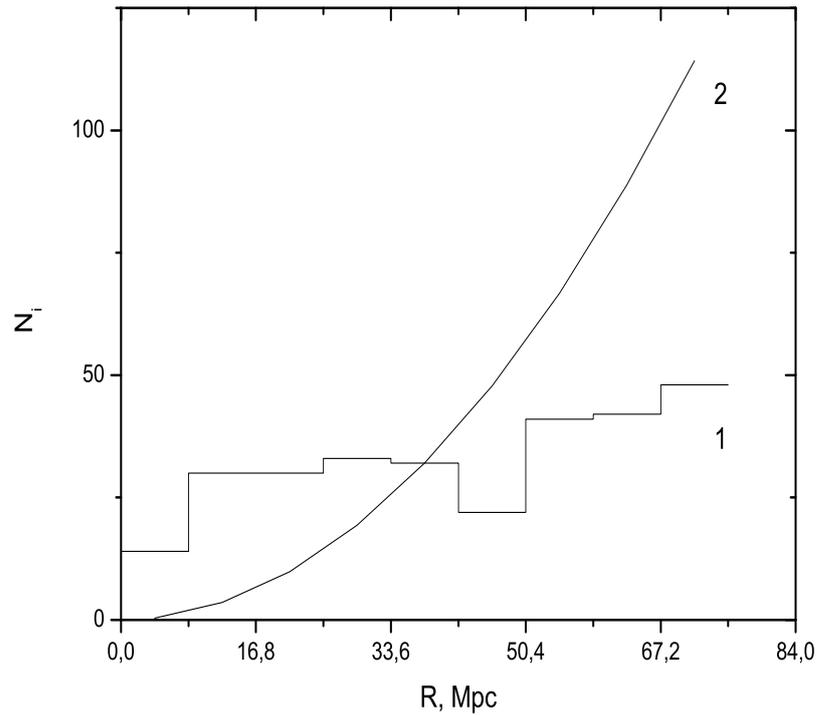

Figure 1. Distribution of the AGN with distance R. 1 – from [4], 2 – in case of isotropic distribution of matter.

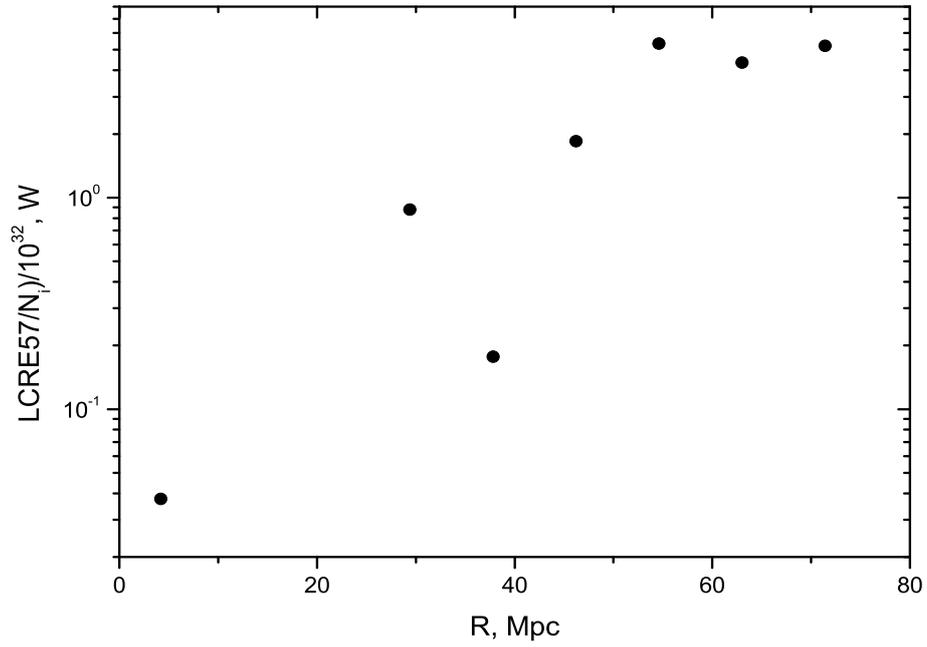

Figure 2. The mean ultrahigh energy cosmic ray luminosity LCRE57 vs. distance R.

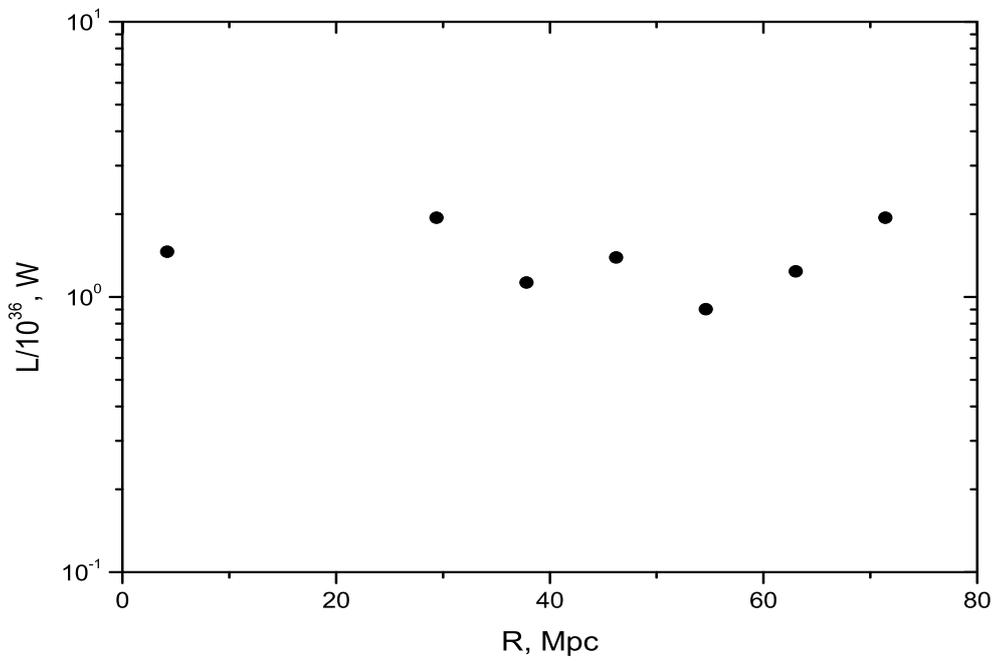

Figure 3. The optical luminosity of the AGN supposed to be the cosmic ray sources vs. distance R.
5

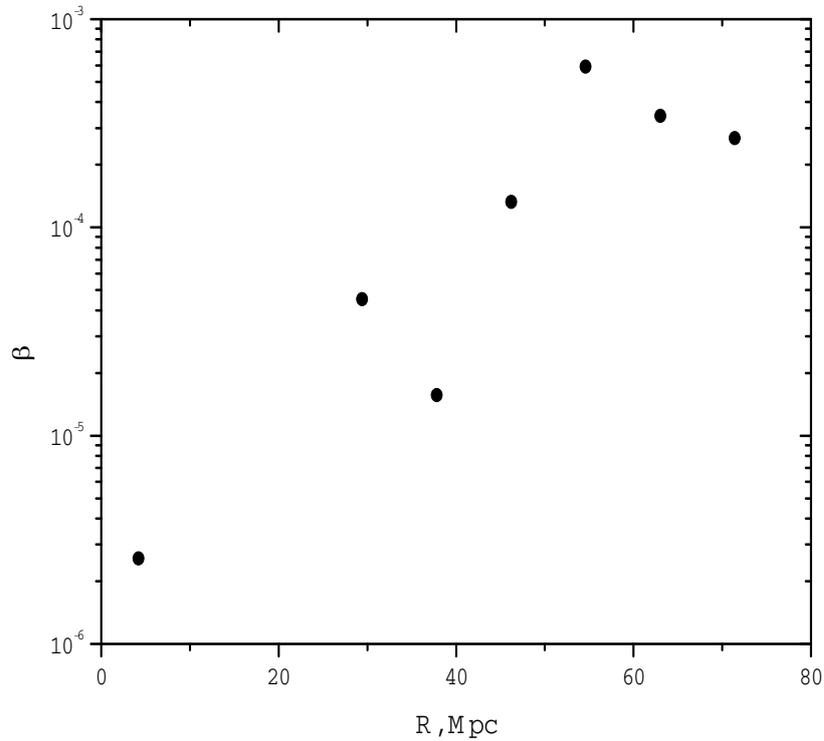

Figure 4. The ratio of cosmic ray luminosity LCRE57 to the optical one L for the AGN supposed to be the cosmic ray sources.

The Fig. 5 displays the normalized number α of cosmic ray sources where α is defined as

$$\alpha = \frac{n_i / N_i}{\sum_{i=1}^{9} n_i / N_i} \qquad (2)$$

in case of observed sources (open circles) and

$$\alpha = \frac{(n_i/N_i)/r_i^2}{\sum_{i=1}^{9}(n_i/N_i)/r_i^2} \qquad (3)$$

in case of expected sources (full circles). Here $n_i$ is a number of the cosmic ray sources in the bin and $N_i$ is a number of the AGN in the same bin. A coefficient $1/r_i^2$ is due to expected decreasing of cosmic ray luminosity with a distance. A dramatic disagreement is seen. The $\chi^2$ test gives $\chi^2=19$ per one degree of freedom. Again it is not easy to understand why at larger distances we have approximately the same number of cosmic ray sources as in the first bin. For comparison the total cosmic ray luminosity of our Galaxy is estimated as $5\times10^{33}$ W [22] that is also should be confronted with the optical luminosity $2\times10^{37}$ W.

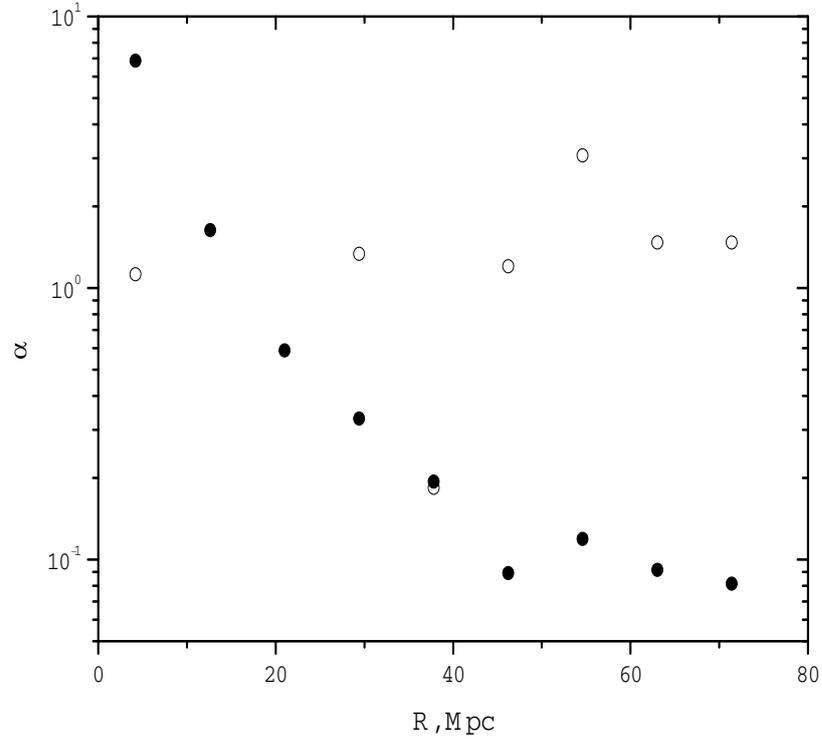

Figure 5. Distribution of normalized numbers of the AGN supposed to be the cosmic ray sources (open circles) and expected numbers (full circles) with distance R.

## 3. Conclusion

So we have to admit that either the nearby AGN emit only cosmic ray particles with energies above ~10 J (~$6\times10^{19}$ eV) or their cosmic ray luminosity considerably exceeds the optical one (the AGN are profound cosmic ray accelerators!). Some alternative way is to assume really nearby sources [23-27]. In this case the GZK suppression of energy spectrum of cosmic rays is not expected. The most valuable contribution to the problem would be construction of the Northern PAO and the Telescope Array (TA) to claim correlations with any objects on the basis of much more larger statistics.

**Acknowledgements.**

Authors thank RFFI (grant 07-02-01212) and the LSS (grant 959.2008.2) for the financial support.